\documentclass[final,5p,times,twocolumn]{elsarticle}
\usepackage{lineno}
\usepackage{amssymb}
\usepackage{algorithm2e}
\journal{Digital Investigation}

\usepackage[utf8]{inputenc}
\usepackage[T1]{fontenc}
\usepackage{tabularx}
\usepackage{placeins}
\usepackage[breaklinks=true]{hyperref}
\usepackage{breakcites}
\usepackage{amsmath}
\usepackage{graphicx}
\usepackage{comment}
\usepackage{xcolor}
\usepackage{float} 

\includecomment{comment} %show comments
\DeclareUnicodeCharacter{2264}{<=}

\begin{document}

%\author{\IEEEauthorblockN{Peter McLaren, Gordon Russell}
%\IEEEauthorblockA{School of Computing,\\
%Edinburgh Napier University,\\
%Edinburgh, UK.}
%\and
%\IEEEauthorblockN{William J Buchanan, Zhiyuan Tan}
%\IEEEauthorblockA{School of Computing,\\
%Edinburgh Napier University,\\
%Edinburgh, UK.}
%}

%\maketitle
\begin{frontmatter}

%% Title, authors and addresses

\title{\bf{Decrypting Live SSH Traffic in Virtual Environments}}

%% use optional labels to link authors explicitly to addresses:
%% \author[label1,label2]{<author name>}
%% \address[label1]{<address>}
%% \address[label2]{<address>}

\author{Peter McLaren, Gordon Russell, William J Buchanan and Zhiyuan Tan}

\address{School of Computing, Edinburgh Napier University, Edinburgh.}

\begin{abstract}
Decrypting and inspecting encrypted malicious communications may assist crime detection and prevention. Access to client or server memory enables the discovery of artefacts required for decrypting secure communications. This paper develops the \textit{MemDecrypt} framework to investigate they discovery of encrypted artefacts in memory and applies the methodology to decrypting the secure communications of virtual machines. For Secure Shell, used for secure remote server management, file transfer, and tunnelling inter alia, \textit{MemDecrypt} experiments rapidly yield AES-encrypted details for a live secure file transfer including remote user credentials, transmitted file name and file contents. Thus, \textit{MemDecrypt} discovers cryptographic artefacts and quickly decrypts live SSH malicious communications including detection and interception of data exfiltration of confidential data. 
\end{abstract}

\begin{keyword}
network traffic; decryption; memory analysis; IoT; Android; VMI; Secure Shell; SSH; AES; Secure File Transfer; data exfiltration; insider attacks;
\end{keyword}

\end{frontmatter}

%% Need to talk about MemCrypt ... framework ... etc earlier on.
%% Can youi make sure you talk about the 'framework'? 
%   PMc DONE: Included in Abstract

% \begin{keywords}
% Keywords—network traffic; decryption; memory analysis; IoT; Android; VMI; Secure Shell; SSH; AES; Secure File Transfer; data exfiltration; insider attacks;
% \end{keywords}

\section{Introduction}
Decrypting malicious communications offers opportunities to discover useful information. This could include botnet command and control traffic identifying compromised machines, confidential information that has been extracted and sent or uploaded to an external location, ransomware keys, or details of criminal activity \cite{Khandelwal}. This paper focuses on decrypting Secure Shell (SSH) traffic, a potential medium for data exfiltration \cite{duncan2015overview}. Realistically useful decryption methods require a knowledge of both the algorithm and the cryptographic artefacts used. Encryption techniques based only on algorithmic secrecy may be unreliable, as mechanisms such as reverse-engineering enable the algorithm’s functionality to be discovered and furthermore, without extensive independent verification, the robustness of an encryption algorithm may be weak \cite{ferguson2011cryptography}. As a result, publicly known encryption algorithms are commonly used, and key secrecy thus becomes paramount. Generating sufficiently long random blocks as keys makes decryption unlikely using brute force methods.

To decrypt, a framework must discover keys and other cryptographic artefacts. When software applications perform encryption and decryption, the artefacts reside in program memory at that moment, whether on the program stack, in the heap, or in shared memory. As memory access is important to forensic investigations \cite{zhang2018memory} software tools and libraries already exist to support such capability for technologies such as desktops, servers, the Internet of Things (IoT), Android smartphones, and virtualized environments. Mechanisms to discover cryptographic artefacts in memory in a manner that allows the target device to continue to operate normally during an investigation while remaining undetectable is of particular interest. This paper presents the \textit{MemDecrypt} framework that stealthily decrypts secure communications traffic. Although earlier researchers have discovered encryption keys in device memory, other cryptographic artefacts, commonly required to decrypt secure traffic, are not considered. \textit{MemDecrypt} implements a novel approach to decrypting SSH traffic by analyzing target memory extracts to identify these candidate cryptographic artefacts (initialization vectors) that, in turn, enable rapid location of candidate keys and the deciphering of payloads in live sessions with high probability. This enables malicious SSH activity in live secure communications sessions to be addressed. The techniques proposed are applicable to a range of device platforms, though the \textit{MemDecrypt} framework is particularly focused on decrypting communications from within virtual machines.

Although plaintext could be obtained by adding an audit function to the binary, this is arguably a different application and has some similarity with a key logger, which may only be acceptable in specific environments. Also, unless all plaintext is captured rather than client input, file contents are not obtained.

Plaintext could possibly be obtained by extracting on buffer memory writes. However, researchers have found that monitoring virtual machine read/write buffers is inefficient. As memory extraction is invasive minimizing the number of extracts is preferable so with buffer memory write triggers, the larger the exfiltrated file, the more extracts. To discover the plaintext of a full session, buffer breaks would need to be in place before the session. In MemDecrypt, memory can be extracted at any stage after the handshake completes to decrypt a captured  network session. Buffer memory write triggers may be effective with interactive sessions as with exfiltrated data, missing an extract makes  decryption problematic. Furthermore, exfiltrating non-ASCII data may be more challenging without certainty of buffer memory locations.

The rest of the paper is structured as follows. To provide framework context, the background to secure communications is provided in Section II. Earlier research in discovering cryptographic artefacts is reviewed in Section III. Section IV presents the \textit{MemDecrypt} design and Section V the implementation details. Test results are evaluated and discussed in Section VI and conclusions drawn in Section VII.

\section{Related Work}

This section provides a summary of symmetric encryption including block and stream algorithms commonly used in secure communications protocols. Approaches for accessing memory to support cryptographic artefact discovery are also discussed.

Although there is no published research into finding cryptographic artefacts in Android smartphone and IOT device memory, desktop and server memory has been studied. Entropy measures have frequently been used as a filtering mechanism to discover keys. This approach assisted in searches for AES key schedules after cold-boot attacks \cite{halderman2009lest} as well in finding Skipjack and Twofish algorithm artefacts \cite{maartmann2009persistence}. These studies focus on encryption key discovery in dormant devices and therefore do not decrypt the secure network sessions of live virtual machines.

Although malware analysis and detection has been a research focus for monitoring from outside the virtual machine, it has also been applied to analyze secure communications. For example, SSH session details were obtained from an SSH honeypot server customized to extract data when the specific system calls executed \cite{sentanoe2017virtual}. In \textit{TLSkex} \cite{taubmann2016tlskex}, AES-CBC cryptographic keys were discovered in Linux client virtual machine memory when Change Cipher Spec messages were detected in TLS network sessions by searching for bit strings where the counts of 0’s and 1’s suggested randomness. \textit{TLSkex} investigates TLS traffic only so, for example, the uploading of confidential data using SSH is not considered. Furthermore, \textit{TLSkex} analysis is restricted to Linux virtual machine so Windows virtual machine activity is excluded. The \textit{MemDecrypt} framework decrypts entire sessions for both SSH and TLS protocols where different encryption algorithms have been applied for Windows clients and Linux servers using a standard entropy measure. Moreover, \textit{MemDecrypt} memory extractions are independent of message type and discovery of candidate initialization vectors drives the decryption process.

Encryption keys can be discovered by intercepting encryption function calls to extract parameters. For example, the Linux \textit{ptrace} command can attach to the encrypting process enabling identification of keys and other artefacts \cite{nakano2014key}. This approach may have been used to discover SSH plaintext, ciphertext, and keys, although implementation details are unclear \cite{hay2012circumventing}. These approaches are Linux-specific and are easily detectable by virtual machine software. Consequently, they may not be effective against malicious insiders, especially when the target device runs Windows. \textit{Ptrace} or the related \textit{strace} can also monitor server system calls to extract SSH plaintext although this presumes control of the server and rapidity of tracing both of which may be problems in live scenarios. \textit{MemDecrypt} decrypts SSH network sessions in a stealthy manner by triggering memory extracts only when an unusual event is detected.

\subsection{Encryption algorithms}
Encryption algorithms for secure communications are asymmetric or symmetric. For encryption and decryption, asymmetric algorithms use different keys whereas symmetric algorithms use the same keys. Asymmetric algorithms attain security through computational complexity, which takes processor time, making them considerably less CPU efficient than symmetric algorithms \cite{puthal2017dynamic}. Consequently, asymmetric algorithms are frequently only used for agreement on symmetric keys, which are then used to encrypt the channel. Symmetric encryption algorithms are either stream algorithms, where plaintext is encrypted with either bit-by-bit or block algorithms (where blocks of a specific size are encrypted). Although the Advanced Encryption Standard (AES) block algorithm may be the \emph{gold standard}, vulnerability and performance concerns have led to the adoption of ChaCha20 stream algorithm with Poly-1305 authentication \cite{nir2018chacha20} in secure protocols such as OpenSSH and OpenSSL, as well as being used for Google Chrome related communications on Android smartphones \cite{Ianix}.

Block and stream algorithms commonly require initialization vectors (IVs) for secure communications. For  AES, IVs incorporated in the encryption process provide defenses against replay attacks \cite{aura1997strategies}. For example, in AES counter mode (AES-CTR), an IV is encrypted and XORed with the plaintext to produce ciphertext. AES-CTR is the quickest AES mode, and is recommended by security experts \cite{ferguson2011cryptography} \cite{rogaway2011evaluation}. For ChaCha20, the key, IV, and a counter are parameters to keystream creation \cite{nir2018chacha20}. The keystream is XORed with the plaintext to produce ciphertext. Both AES-CTR and ChaCha20 are approved for SSH \cite{bellare2005secure} and TLS protocols. Consequently, encryption keys and IVs must be discovered to decrypt AES-CTR and ChaCha20 encrypted SSH and TLS channels.

This paper focuses on SSH communications. For SSH in AES-CTR mode, the IV increments by 1 for each outgoing plaintext block \cite{dworkin2001recommendation} so that the difference between the IV for the first plaintext block in packets n+1 and n is the number of plaintext blocks in packet n. Although AES-CTR is the only recommended SSH AES mode \cite{bellare2005secure}, AES-CBC is also used.  For AES-CBC, each IV after the initial value is the ciphertext of the previous block \cite{dworkin2001recommendation}. Consequently, the IV for each encrypted AES-CBC block is known. ChaCha20 uses the IV to generate key streams. It performs 20 rounds of mathematical operations starting from a base structure consisting of a constant string of 16 bytes, a generated 32-byte key, a 4-byte counter, and a 12-byte IV, where the counter is typically 0 or 1 for each 64-byte plaintext block \cite{nir2018chacha20}. 

\begin{figure*}[!htb]
        \center{\includegraphics[width=0.9\textwidth]
        {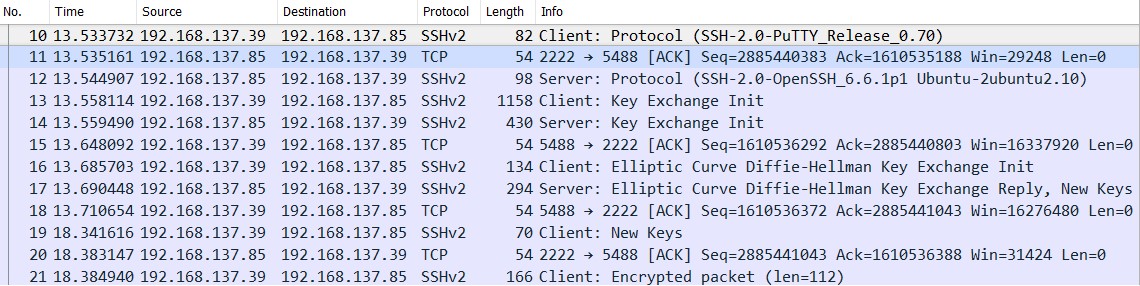}}
        \caption{\label{fig:SSHhandshake} SSH Handshake Example}
      \end{figure*}

SSH enables secure management of remote servers across potentially insecure networks, offering functionality such as client-server file transfer. The protocol is specified in 4 key IETF RFCs: SSH Protocol Architecture (SSH-ARCH) \cite{ylonen2005arch}, SSH Transport Layer Protocol (SSH-TRANS)\cite{ylonen2005transport}, SSH Authentication Protocol (SSH-AUTH) \cite{ylonen2005auth}, and the SSH Connection Protocol (SSH-CONNECT) \cite{ylonen2005connection}. SSH-TRANS defines the initial connection, packet protocol, server authentication, and the basic encryption and integrity service \cite{barrett2001ssh}. Following the TCP handshake, the parties transmit supported SSH protocol versions, and optionally application, which enables the probable operating systems and library to be inferred. For instance, 'SSH-2.0-PuTTY\_Release\_0.70’ probably signifies that a Windows client is executing the PuTTY application \cite{PuTTY}. Exchanged 'Key Exchange Initialization; and 'Key Exchange' messages determine the session encryption and authentication algorithm and the material for the generation of the cryptographic artefacts. Client \emph{New Keys} messages advises that all subsequent traffic in the session is encrypted. An example of the handshake process as well as the first encrypted packet is illustrated in Figure \ref{fig:SSHhandshake}. SSH-AUTH defines authentication methods such as \emph{public key}, \emph{password}, \emph{host based} and \emph{none}.  After successful authentication, a file transfer requires the establishment of a secure channel to support the secure file transfer protocol as defined by. Secure file transfer (SFTP) \cite{galbraith2006ssh} is an SSH sub-system particularly worthy for investigation as significant potential exists for it to transfer confidential files out of a system. 
 
\subsection{Memory Access}
Memory acquisition tools assist forensic analysis. So, for workstation and server technologies hardware and software acquisition methods exist \cite{vomel2013evaluation}. Hardware acquisition typically involves connecting devices, such as PCMCIA cards or USB sticks, to a target \cite{vomel2011survey} while software acquisition commonly involves executing extraction programs such as FTK Imager \cite{FTKImager}, Memoryze \cite{Memoryze}, or WinPmem \cite{WinPMEM} on the target \cite{freiling2018advances}. These solutions may not always be practical in live network session decryption scenarios. 

Android smartphone volatile memory is accessible. As Androids run Linux, memory acquisition tools such as the Linux Memory Extractor ('LiME’) application \cite{LiME} may suffice. However, LiME depends on compiled kernel modules for the target’s Linux version, support by the smartphone and kernel level execution. The quantity of Linux variations for Android smartphones as well as the installation and execution requirements may be challenging. AMExtractor \cite{yang2016tool}  requires kernel execution privilege but no compilation is required and so is potentially less restrictive. TrustDump \cite{sun2015reliable} may be appropriate but minimal testing has been carried out.   Commercial tools such as Cellebrite also claim to extract memory from Android devices without target modification although usage is restricted \cite{Cellebrite}. 

Internet of Things (IoT) devices also commonly run Linux \cite{case2017memory}. However, device type and Linux variations pose potentially greater challenges than smartphones. Nevertheless, solutions that support live acquisition from Android smartwatches, as well as smartphones, have been proposed \cite{yang2017live}. IoT device memory may also be acquired by flashing memory, running Linux dump commands, or accessing device circuitry \cite{kondapally2018}. Furthermore, memory access with commercial tools, such as Cellebrite UFED Physical Analyzer, has also been demonstrated \cite{alabdulsalam2018internet}. As IoT devices frequently communicate with cloud-based servers, memory acquisition of virtualized machines may present an easier alternative \cite{case2017memory}.
Virtualization enables memory access. Virtualization technologies enable virtual machines to share host computer resources thereby providing an opportunity to discover cryptographic artefacts in virtual machine memory from the physical host. This ensures investigations have reduced the impact on virtual machine operations. Furthermore, software programs executing on the virtual machine, such as malware, may not detect the investigations. Examples of tools and libraries that support outside\-the-machine monitoring include LibVMI \cite{LibVMI} together with PyVMI \cite{pyvmi} and Volatility \cite{volatility2016}, and Rekall \cite{rekall2017}.

\section{MemDecrypt Design}
\begin{figure*}[!htb]
        \center{\includegraphics[width=0.7\textwidth]
        {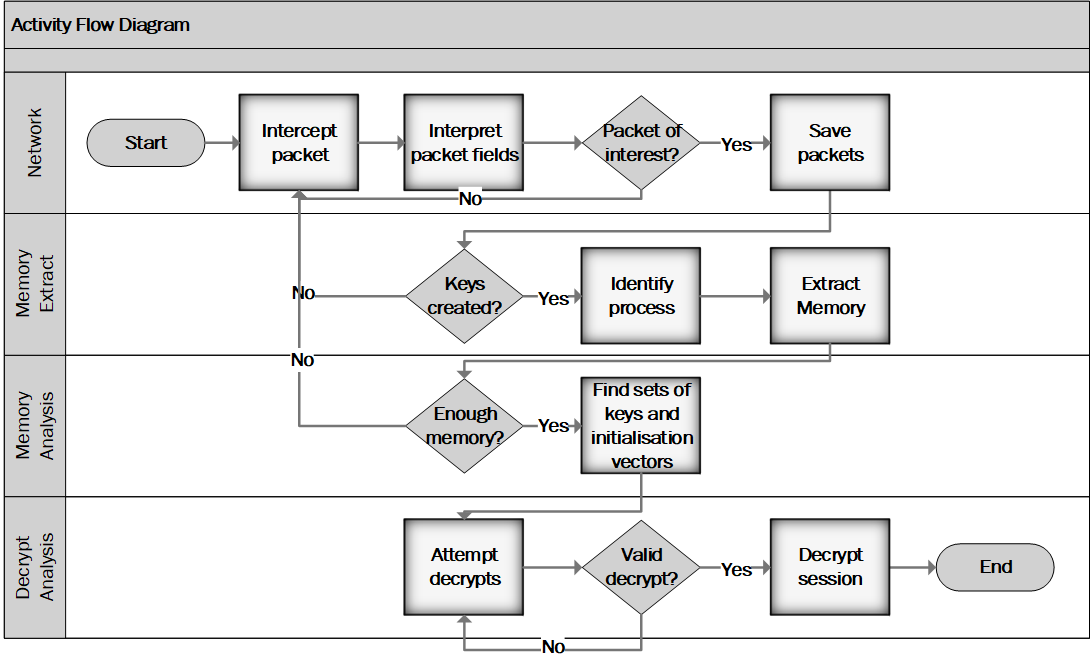}}
        \caption{\label{fig:ActivityFlow} \textit{MemDecrypt} Activity Flow Diagram}
      \end{figure*}

%% Need to mention in title. PMc: The name MemDecrypt?
\textit{MemDecrypt} consists of network and data collection, memory analysis, and decrypt analysis components. Figure \ref{fig:ActivityFlow} illustrates the \textit{MemDecrypt} activity flow diagram. Each component is described in the following paragraphs. 

\textbf{Network and Memory Extract}. In \textit{MemDecrypt} unusual events trigger memory extracts. This approach is less intrusive than continuous memory monitoring where the monitoring and analysis activities of the host may impact target device performance. Furthermore, malware writers script programs to be aware of monitoring activity, which would probably be more obvious with continuous monitoring. The triggers approach is also more precise than obtaining memory snapshots on a polled basis. Polling snapshots may miss malicious activity if the polling interval is too large, especially when malware uses counter analysis techniques. The quantity and timing of memory extraction events depend on the target device, the secure protocol, and the encryption algorithm. Where memory is classifiable, the read/write memory of the encryption program is extracted for size minimization, with consequent reduced impact on target performance and faster subsequent analysis.

\textbf{Memory analysis}. Candidate encryption keys and IVs are identified in the memory extracts. Although largely protocol specific, there are common features. In particular, candidate IV locations are discovered first with approaches that encompass an analysis of memory extracts, network packets or both network packets and memory extracts. As keys and IVs are cryptographic artefacts, the distance between their respective memory locations may be small If program memory extracts are taken when the same activity is being performed, such as the transmission of outgoing messages, memory blocks containing IVs change, while other blocks remain static. 

Key randomness makes it different from many other types of memory regions. Key randomness means that the sequence of bits cannot be easily predicted. The randomness of keys can be evaluated using entropy, a measure of the amount of information in a key. This paper uses Shannon’s entropy measure for discrete variables \cite{shannon1948mathematical} in preference to cryptographically useful alternatives such as guessing entropy and min-entropy \cite{cachin1997entropy} because smaller candidate key sets are produced:
\begin{equation}
H=-\sum_{i=1}^{n} p(i) \log_2 p(i)
\end{equation}
where $p(i)$ is the normalized frequency of the $i$th byte in the message i.e. $p(i) = f(i)/n$. So, segments of high entropy user memory are more likely to contain the key. In contrast with IVs, keys do not generally change during a session. So, static, high-entropy contents are candidate encryption keys. This observation assists in improving memory analysis performance.

\textbf{Decrypt analysis}. Candidate keys and IVs identified in memory analysis are used in decrypting network packets until a valid key and IV combination has been found. Decrypt validation uses information derived from specific encrypted fields. SSH encrypted data blocks are of the following format:

Packet Length 
(4 bytes)    Padding Length  
(1 byte)    Payload
(variable bytes)    Padding
(variable bytes)    MAC

The packet length is the sum of the padding length size, the payload, and padding fields. So, equation (2) is a good decrypt test for many SSH messages as $2^{(8*4-21)}$ valid packet length decrypts are possible. The minimum SSH block size is 21 bytes comprising a packet length of 4 bytes, a padding length of 1 byte, and the payload and padding which is at least one block. So, the probability of an incorrect decrypt producing the correct header data is 1-in-4,294,967,275. Reassembly is undertaken when the SSH packet size exceeds the network packet size. Equation (2) is sound during the authentication, channel, and sub-service setup stages when SSH packet sizes are generally small and a modified version is used for reassembled SSH packets. An additional test evaluates whether the decrypted padding length meets Equation (3) as required by SSH-TRANS. Correct decrypts are parsed to obtain SSH and SFTP fields.

\begin{equation}
\begin{split}
packet\ data\ length\ = \\ 
     & decrypted\ packet\ length\ + \\
     & size(packet\ length\ field)\ + \\
     & size(MAC field)
\end{split}
\end{equation}
\begin{equation}
       4 ≤ padding\ length ≤ 255  
\end{equation}
\section{MemDecrypt Implementation}
This paper focuses on SSH decryption using AES-CTR and AES-CBC in virtualized environments using \textit{MemDecrypt}. The following paragraphs present implementation and evaluation details. The framework is implemented on the Xen hypervisor \cite{xen2018}. Xen’s small trusted computing base makes it potentially less prone to vulnerabilities than hypervisors with larger footprints. Furthermore, the LibVMI library (“LibVMI,” n.d.) for Xen enables efficient memory access to live memory of Windows or Linux virtual machines. As the Xen hypervisor has minimal functionality a privileged virtual machine (Dom0) manages the hypervisor and provides network and virtual disk device access to other virtual machines. Network access for the virtual machines is through a Dom0 virtual software bridge. The \textit{MemDecrypt} components either all run on, or are initiated, from Dom0. 

\begin{figure*}[!htb]
        \center{\includegraphics[width=0.7\textwidth]
        {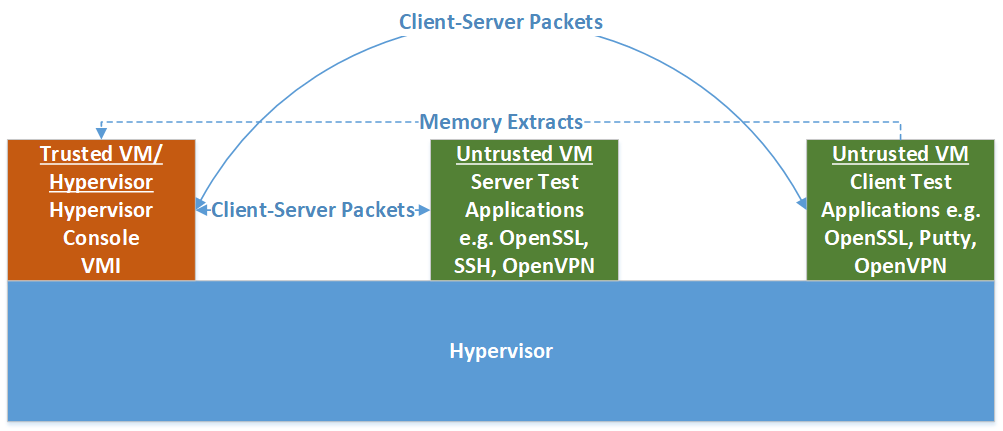}}
        \caption{\label{fig:Architecture} \textit{MemDecrypt} Virtualization Architecture}
      \end{figure*}

The \textit{MemDecrypt} implementation architecture for virtualized environments is illustrated in Figure \ref{fig:Architecture}. An isolated hypervisor supports two unprivileged virtual machines, shown in the centre and right of the figure, and one privileged virtual machine shown on the left. Test client applications execute on the virtual machine on the right, targeting server applications executing on the virtual machine, shown in the centre. 
 
\subsection{Data Collection}
%% Title doesn't look right? PMc: Its seems OK to me
For virtualized environments, virtual machine network traffic is inspected by redirecting each packet to a local queue using an iptables rule and NetFilterQueue 0.8.1 \cite{netfilterqueue}, and analyzing protocol fields using Scapy 2.3 \cite{scapy2017}. When unusual activity is detected, the component stores the network packet and deconstructs the message. Memory is extracted for any 2 outgoing SSH messages after a \emph{New Keys} message. Linux memory extraction uses PyVMI and LibVMI libraries, whereas Windows extraction applies Volatility framework user plugins.

\textit{MemDecrypt} obtains useful data from the SSH initialization stage. Client and server versions, and application if available are obtained from the protocol version exchange. The encryption algorithm is determined from the “Key Exchange” messages. Also, if initialization has completed, i.e. the “New Keys” has been transmitted, user-level read/write program memory extraction is triggered for two outgoing packets in the network session. Memory extracts are not required for consecutive packets or to be immediately after the “New Keys” message.  

\subsection{Memory Analysis}
Analysis approaches vary according to encryption mode and operating system. For AES-CTR, two steps are required to discover candidate IVs and keys in memory, whereas AES-CBC requires only key discovery. For Windows, discovery is performed by iteratively analyzing multiple memory files extracted at different times, whereas, for Linux, a single heap file is analyzed.

For AES-CTR, candidate IVs are discovered first. As IVs increase but are likely to be located at the same memory address over different extracts, memory blocks that change is subject to further analysis. If the 16-byte value at a memory address increments by the number of encrypted blocks in the previous packet, then the address contents are a candidate IV. Supposing that value at location $p$ in capture $y$ at the time $a$ is compared with the value at location $p$ in capture $y$ at time $b$. Then, if the values are IVs and represented by $IV_{pya}$ and  $IV_{pyb}$ respectively, then $IV_{pyb} = IV_{pya}+n$, where $n$ is the number of AES encrypted network blocks that have been sent between the time $a$ and $b$ in that session. For example, if the value of a 16-byte memory block is 123456 and two network packets with, say, 10 and 5 encrypted blocks are sent and captured, then a value of 123471 at the same position in the later extract identifies a candidate AES-CTR IV. Algorithm \ref{alg:AES-CTRIV} shows the process.

\begin{algorithm}
\SetAlgoLined
\KwData{extract folders $fldr_a$, $fldr_b$ and packets $pkt_a$, $pkt_b$}
\KwResult{Z = candidate IVs}
delta := blocks[$pkt_a$:$pkt_b$]\;
\For {file $f_1$ in $fldr_a$} {
    $f_2$ = match ($f_1$, $fldr_b$)\;
    \If{$f_1$ <> $f_2$}{
        \For{i = 0 to size($f_1$) inc 4}{
            \If {val($f_2$[i:i+16]) - val($f_1$[i:i+16]) = delta}{
                Z += $f_1$[i:i+16]\;
            }
        }
    }
}
\label{alg:AES-CTRIV}
\caption{AES-CTR IV Memory Analysis}
\end{algorithm}

To discover AES candidate keys for AES-CTR and AES-CBC, the memory extract files are analysed. Key segment entropies are calculated for key length segment sizes. If an entropy exceeds a threshold, the segment is compared with the equivalent segment in a later extract, and if the segments are identical, the segment is a candidate encryption key. For example, a 256-bit key length, a 32-byte memory segment entropy of 4.9, and a 32-byte AES threshold of 4.65 determines the segment to be of interest. An identical match to the segment at the same location in a later memory extract identifies a candidate key. The identified candidate IVs and keys provide input to the decrypt analysis stage. Heuristic testing determined that AES entropy thresholds of 4.65 for 256-bit keys, 4.0 for 192-bit keys, and 3.4 for 128-bit keys ensured the inclusion of all keys in candidate sets while minimizing set size. 

\subsection{Decrypt Analysis}
The component iterates through each candidate key for each candidate IV until decrypts are validated. The first ciphertext block is decrypted for each combination with pycrypto 2.6.1 \cite{pycrypto2018}. For a correct decrypt the first four plaintext bytes are the packet length and Equation (2) holds. For additional validation, the decrypted padding length is checked with Equation (3).  With a valid key and IVs, \textit{MemDecrypt} decrypts each block and deconstructs the SSH plaintext stream. For SSH authorization requests, the 'password’ type plaintext yields the remote user credentials and for SSH connection requests, the channel type, and channel request decrypts. For SFTP, all plaintext is produced including \emph{initialization}, \emph{file attribute}, \emph{file open}, \emph{write} and \emph{close} message types fields. All plaintext is written to file for evaluation. 

\subsection{Testbed}
%% Title? PMc: Sub-section was intended
The physical environment is a Core 2 Duo Dell personal computer with 40\,GB of disk storage and 3\,GB of RAM. It hosts the hypervisor, a Dom0 privileged virtual machine, an untrusted Windows virtual machine, and an untrusted Ubuntu virtual machine. The hypervisor is Xen Project 4.4.1 and the Dom0 hypervisor console is Debian release 3.16.0-4-amd64 version 1. 

Tests run on Windows client and Linux server virtual machines. One client runs a standard Windows 7 SP1 operating system with 512\,MB of allocated memory and 30\,GB of disk space. Another client runs a Windows 10 (10.0.16299) operating system with 2 GB of memory and 40\,GB of disk. Windows operating systems support a number of SSH clients \cite{ssh2018}. The selected PuTTY suite \cite{PuTTY} is widely used \cite{ssh2018} so may be used by suspect actors. However, other SSH client applications should produce similar results. 
The untrusted Linux server virtual machine runs an Ubuntu 14.04 build (“Trusty”) with 512\,MB of allocated memory and 4\,GB of disk storage. SSH server functionality is provided by openssh-server. To remove unnecessary communications with external agents, the \emph{dnsmasq} package is installed and configured to respond to DNS requests with the server virtual machine IP address.

\section{Evaluation}
\textit{MemDecrypt} is evaluated by running a sequence of experiments. The experimental set-up is described followed by the presentation and review of results. Possible countermeasures to \textit{MemDecrypt} results are discussed.
\subsection{Experimental Set-up}
Experiments are performed with variable file sizes, key lengths, modes of operation, operating systems, and operating system versions. In each instance, the 'pscp’ program is executed from the Windows command line using requests of the form: \\

\textit{pscp  -P  nnnn  filename  name@ipaddress:/home/name} \\

where \textit{nnnn} is the target port, \textit{filename} is the file being transmitted, \textit{name} is a user account on the target Ubuntu server, \textit{ipaddress} is the target server IP address and \textit{/home/name} is the Ubuntu server target folder for the transmitted file. An Ubuntu service is started from the bash command line to listen to client SSH messages with requests of the form:\newline

\textit{/usr/sbin/sshd -f /root/sshd\_config -d -p nnnn}\newline

where \textit{nnnn} is the service receiving port number and \textit{sshd\_config} contains configuration details such as encryption algorithms supported by the server. 

Sets of experiments investigate decrypting SSH traffic encrypted with AES under different conditions. One set evaluates decrypt effectiveness for Windows 7 and Windows 10 clients. A second set evaluates the effectiveness of 128-bit, 192-bit and 256-bit keys on Windows 10 clients in AES-CTR mode. A third set evaluates \textit{\textit{MemDecrypt}} effectiveness with 256-bit keys in AES-CBC and AES-CTR modes on Windows 10 clients. To evaluate file invariability, a fourth set uploads 30 files in text, pdf, Excel, and executable formats between 1\,KB and 500\,KB for Windows clients in AES-CTR mode using 256-bit keys. Experiments also assess decrypt effectiveness with Ubuntu server for AES-CBC and AES-CTR with 256-bit keys. 

\subsection{Test Results}
\begin{figure*}[!ht]
        \center{\includegraphics[width=0.75\textwidth]
        {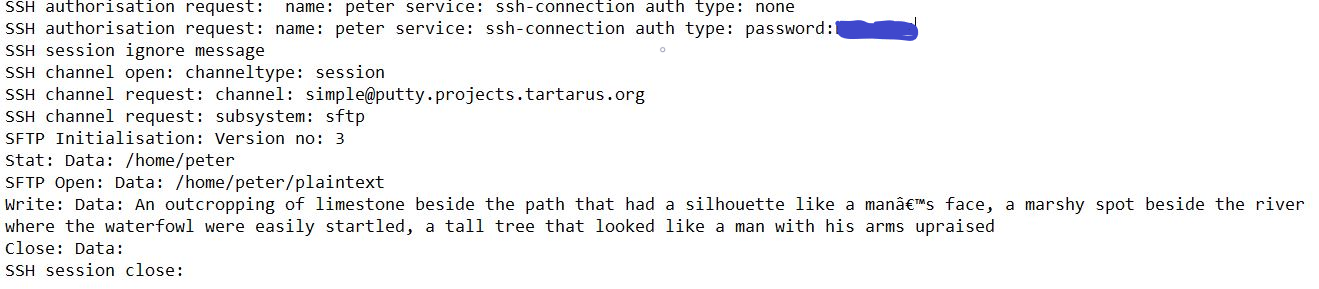}}
        \caption{\label{fig:SSHdecrypt}SSH Decrypt Output}
      \end{figure*}

In each experiment, encryption keys, and for AES-CTR initialization vectors, were discovered and valid plaintext produced for all SSH and SFTP fields. For example, with a client command of 'pscp -P 2222 plaintext.txt peter@192.168.137.85:/home/peter’ and plaintext.txt of 'An outcropping of limestone beside the path that had a silhouette…’, the interesting decrypted fields are depicted in Figure \ref{fig:SSHdecrypt}. \textit{MemDecrypt} also produces other SSH fields such as request identifiers, and file offsets. As observed earlier, the probability of an incorrect combination generating a packet length meeting Equation (2) is 0.00000002\% (1 in 4,294,967,275). \textit{MemDecrypt} decrypts SSH traffic with a high degree of certainty.
 
Analysis durations for producing correct plaintext determines \textit{MemDecrypt}’s usefulness. For example, if plaintext is produced during the network session \textit{MemDecrypt} can assist in the prevention of further malicious activity, perhaps by dropping packets or hijacking the session.

The first experiment compares the relative performance of Windows 7 and Windows 10 client virtual machines. For AES-CTR, two memory extracts are required for the analysis whereas, for CBC, one extract suffices. Memory analysis typically executes for approximately nine seconds for Windows 7 clients and 16 seconds for Windows 10 clients with a maximum of 25.1 seconds. Decrypt analysis durations varied between 0.2 and 34.1 seconds averaging at 4.5 seconds. The variance is linked to the candidate IV set size and the ordinality of the correct IV within the file set.

\begin{figure*}[!htb]
        \center{\includegraphics[width=0.5\textwidth]
        {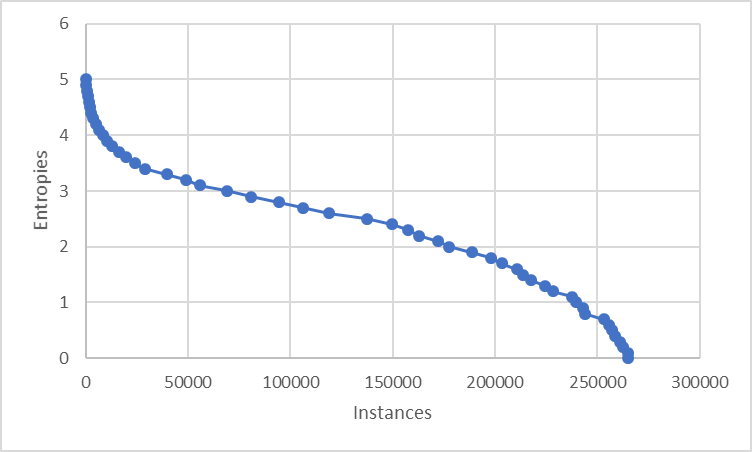}}
        \caption{\label{fig:EntropyDistrib}Typical Memory Segment Entropy Distribution}
      \end{figure*}

\begin{figure*}[htb!]
        \center{\includegraphics[width=0.45\textwidth]
        {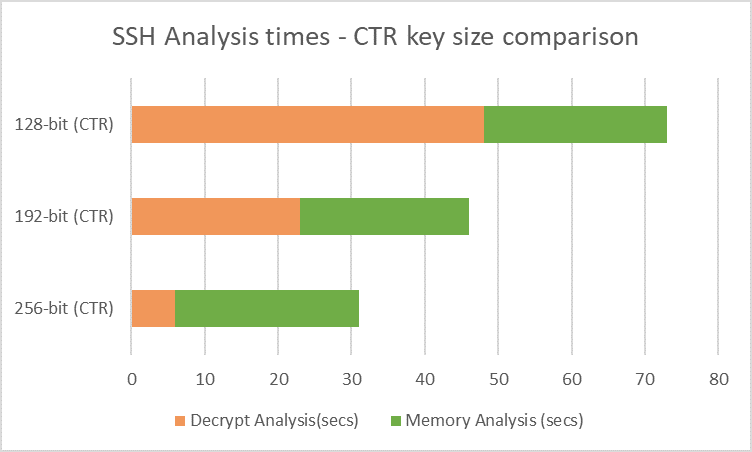}}
        \caption{\label{fig:Durations} Key Length Analysis Durations}
      \end{figure*}

The second experiment compares analysis time durations for different CTR key sizes on Windows 10 clients. Shorter key lengths require lower entropy thresholds, so more candidate encryption keys are discovered in-memory analysis. Figure \ref{fig:EntropyDistrib} illustrates a typical distribution of 32-byte entropy segments in read/write memory. This maps the count of memory segments exceeding an entropy with an entropy levels so that for example whereas out of 264,813 segments exceeding 0.0 entropy, 188,602 (i.e. 72.1\%) exceed 2.0, 2,628 (i.e. 0.99\%) exceed 4.5. So, for example, in one test sequence memory analysis yielded candidate key set sizes of 272 for 256-bit key lengths, 1123 for 192-bit key lengths, and 5658 for 128-bit key lengths. With these set sizes, decrypt analysis durations are longer for shorter key lengths as illustrated in Figure \ref{fig:Durations}.

The third experiment compares analysis time durations on Windows 10 clients for 256-bit key sizes in AES-CTR and AES-CBC. The CBC memory analysis takes approximately 16 seconds which is similar to CTR. However, the CBC decrypt analysis duration is faster with a minimum of 0.07 seconds as iterating through potential IVs is not required. 
 
For experiments accessing Ubuntu server memory with the default encryption algorithm, i.e. AES with 256-bit key length and CTR mode, all client and server packets are correctly decrypted. The data collection component obtains process lists and extracts process heap from the Ubuntu virtual machine in 0.3 seconds. Memory analysis  finds approximately 320 keys and 3 initialization vectors in 6 seconds, and  decrypt analysis  decrypts the session successfully in 37 seconds.

\textit{MemDecrypt} performance may suffice when extracts are obtained for Windows clients or Ubuntu servers. Nevertheless, strategies to enhance performance include improving memory extraction for Windows clients, pre-testing with known SSH client and server applications, pipelining, multi-threading, and implementing in a low-level language instead of Python. A custom extract engine using PyVMI and LibVMI libraries to replace Volatility plugins improves Windows memory extraction performance. Pre-testing SSH client and server applications may determine the distance between key and IV memory locations. 

Cryptographic libraries generally request memory to hold crypto data structures ('malloc') when algorithms are agreed which occurs after the handshake so data is usually on the heap. The data structures can include fields such as encryption/decryption flag, key size, keys etc so for an algorithm, AES-CTR with 256 bit keys, the data structures may be invariant. For example, with PuTTY 'pscp’,  distances are 968 bytes for 256-bit and 192-bit keys and 728 bytes for 128-bit-keys and are invariant with operating system version or transmitted file size. Where the distance is known, and the program identified from the SSH version message, memory analysis and decrypt analysis components take one second. Multi-threading supports simultaneous analysis of multiple files and decrypts while pipelining between components enables analysis  to terminate when the correct plaintext is obtained.

So, \textit{MemDecrypt} decrypts SSH sessions with high probability independent of file size, operating system type or version, key length, or mode. Furthermore, with SSH application pre-testing, analysis and decrypt decryption completes in 1 second. With unknown SSH applications, the plaintext is produced in under 60 seconds for 192-bit and 256-bit keys. Although in experiments, MemDecrypt decrypts sessions with exfiltrated files of 100 bytes, the risk exists that extracts are not acquired in terse SSH sessions. The risk might be mitigated by pausing the virtual machine. Decrypting sessions with SSH key rotation \cite{ylonen2015security} is not currently implemented but the planned MemDecrypt approach is considering each rotation as a separate session with its own candidate keys and IVs. 

%\begin{figure*}[!htb]
%        \center{\includegraphics[width=0.4\textwidth]
%        {figures/SSHMessageFlow.png}}
%        \caption{\label{fig:SSHMessageFlow} SSH Message Flow}
%      \end{figure*}

\subsection{Countermeasures}
Countermeasures may prevent or delay \textit{MemDecrypt} discovery of cryptographic artefacts. Invalid assumptions can invalidate the methodology. Candidate encryption keys are assumed to be high entropy, static for a network session, and in the same memory location. For entropy, less randomness, i.e. lower entropy, makes key regions less evident but key unpredictability is an essential requirement. For key staticity, \textit{MemDecrypt} requires two extractions for AES-CTR, key changes would be required between each outgoing packet which could cause excessive transmission delays. Key location changes could delay decryption. However, tests on a Linux heap extract produced delays of less than 0.5 seconds.  \textit{MemDecrypt} assumes candidate AES-CTR IVs are located at the same memory locations in each extract and values to increment by the sum of payload blocks in the previous packets.  As with keys, tests where IV memory addresses changed induced delay of 0.5 seconds. As a result, the measure may not suffice. AES-CTR IVs increments make them detectable when stored \emph{in the clear} in memory. Another delaying measure is encrypting artefacts with an additional key. However,  this key may  be discoverable, and furthermore, the additional encryption and decryption for each packet, or block, may have an unacceptable performance impact. Obfuscation  the artefacts may be more effective.  For example, splitting key and IV strings and interpolating variable data between splits will limit \textit{MemDecrypt} performance, and possibly effectiveness. {This technique is faster and less detectable than an additional encryption layer. A more effective counter-measure is preventing memory access to artefacts. For example, Intel \cite{patel2017} and AMD \cite{nichols2018} may develop virtual machine encryption where encryption keys are absent from virtual machine memory. Although this can offer privacy, malicious behaviour is then hidden so administrators may seek to disable the feature.

\section{Conclusions and Future Work}

The \textit{MemDecrypt} framework rapidly discovers cryptographic artefacts and decrypts SSH communications in virtualized environments. This can assist in detecting, and preventing insider attackers from extracting and encrypting confidential information to external locations. \textit{MemDecrypt} can be extended to technologies where memory acquisition of live secure sessions is enabled. Decrypting SSH sessions may be illegal without approval so cryptographic artefact sets could be retained with the associated network traffic for decryption once approval is obtained.  High performance makes the framework applicable so future work should apply multithreading and pipelining techniques before being extended to other non-virtualized use cases, secure protocols, encryption algorithms, and malware that use encrypted communications channels.

\bibliographystyle{IEEEtran}
\bibliography{main}

% Generated by IEEEtran.bst, version: 1.14 (2015/08/26)
\begin{thebibliography}{10}
\providecommand{\url}[1]{#1}
\csname url@samestyle\endcsname
\providecommand{\newblock}{\relax}
\providecommand{\bibinfo}[2]{#2}
\providecommand{\BIBentrySTDinterwordspacing}{\spaceskip=0pt\relax}
\providecommand{\BIBentryALTinterwordstretchfactor}{4}
\providecommand{\BIBentryALTinterwordspacing}{\spaceskip=\fontdimen2\font plus
\BIBentryALTinterwordstretchfactor\fontdimen3\font minus
  \fontdimen4\font\relax}
\providecommand{\BIBforeignlanguage}[2]{{%
\expandafter\ifx\csname l@#1\endcsname\relax
\typeout{** WARNING: IEEEtran.bst: No hyphenation pattern has been}%
\typeout{** loaded for the language `#1'. Using the pattern for}%
\typeout{** the default language instead.}%
\else
\language=\csname l@#1\endcsname
\fi
#2}}
\providecommand{\BIBdecl}{\relax}
\BIBdecl

\bibitem{Khandelwal}
S.~Khandelwal, ``{“How Dutch Police Decrypted BlackBerry PGP Messages For
  Criminal Investigation,” The Hacker News},''
  \url{https://thehackernews.com/2017/03/decrypt-pgp-encryption.html}, 2017,
  [Online; accessed 29-Jan-2019].

\bibitem{duncan2015overview}
A.~Duncan, S.~Creese, and M.~Goldsmith, ``An overview of insider attacks in
  cloud computing,'' \emph{Concurrency and Computation: Practice and
  Experience}, vol.~27, no.~12, pp. 2964--2981, 2015.

\bibitem{ferguson2011cryptography}
N.~Ferguson, B.~Schneier, and T.~Kohno, \emph{Cryptography engineering: design
  principles and practical applications}.\hskip 1em plus 0.5em minus
  0.4em\relax John Wiley \& Sons, 2011.

\bibitem{zhang2018memory}
N.~Zhang, R.~Zhang, K.~Sun, W.~Lou, Y.~T. Hou, and S.~Jajodia, ``Memory
  forensic challenges under misused architectural features,'' \emph{IEEE
  Transactions on Information Forensics and Security}, vol.~13, no.~9, pp.
  2345--2358, 2018.

\bibitem{halderman2009lest}
J.~A. Halderman, S.~D. Schoen, N.~Heninger, W.~Clarkson, W.~Paul, J.~A.
  Calandrino, A.~J. Feldman, J.~Appelbaum, and E.~W. Felten, ``Lest we
  remember: cold-boot attacks on encryption keys,'' \emph{Communications of the
  ACM}, vol.~52, no.~5, pp. 91--98, 2009.

\bibitem{maartmann2009persistence}
C.~Maartmann-Moe, S.~E. Thorkildsen, and A.~{\AA}rnes, ``The persistence of
  memory: Forensic identification and extraction of cryptographic keys,''
  \emph{digital investigation}, vol.~6, pp. S132--S140, 2009.

\bibitem{sentanoe2017virtual}
S.~Sentanoe, B.~Taubmann, and H.~P. Reiser, ``Virtual machine introspection
  based ssh honeypot,'' in \emph{Proceedings of the 4th Workshop on Security in
  Highly Connected IT Systems}.\hskip 1em plus 0.5em minus 0.4em\relax ACM,
  2017, pp. 13--18.

\bibitem{taubmann2016tlskex}
B.~Taubmann, C.~Fr{\"a}drich, D.~Dusold, and H.~P. Reiser, ``Tlskex: Harnessing
  virtual machine introspection for decrypting tls communication,''
  \emph{Digital Investigation}, vol.~16, pp. S114--S123, 2016.

\bibitem{nakano2014key}
Y.~Nakano, A.~Basu, S.~Kiyomoto, and Y.~Miyake, ``Key extraction attack using
  statistical analysis of memory dump data,'' in \emph{International Conference
  on Risks and Security of Internet and Systems}.\hskip 1em plus 0.5em minus
  0.4em\relax Springer, 2014, pp. 239--246.

\bibitem{hay2012circumventing}
B.~Hay and K.~Nance, ``Circumventing cryptography in virtualized
  environments,'' in \emph{Malicious and Unwanted Software (MALWARE), 2012 7th
  International Conference on}.\hskip 1em plus 0.5em minus 0.4em\relax IEEE,
  2012, pp. 32--38.

\bibitem{puthal2017dynamic}
D.~Puthal, S.~Nepal, R.~Ranjan, and J.~Chen, ``A dynamic prime number based
  efficient security mechanism for big sensing data streams,'' \emph{Journal of
  Computer and System Sciences}, vol.~83, no.~1, pp. 22--42, 2017.

\bibitem{nir2018chacha20}
Y.~Nir and A.~Langley, ``Chacha20 and poly1305 for ietf protocols,'' Tech.
  Rep., 2018.

\bibitem{Ianix}
Ianix, ``{“ChaCha Usage \& Deployment,” Ianix},'' \url{https://ianix.com},
  2019, [Online; accessed 29-Jan-2019].

\bibitem{aura1997strategies}
T.~Aura, ``Strategies against replay attacks,'' in \emph{Computer Security
  Foundations Workshop, 1997. Proceedings., 10th}.\hskip 1em plus 0.5em minus
  0.4em\relax IEEE, 1997, pp. 59--68.

\bibitem{rogaway2011evaluation}
P.~Rogaway, ``Evaluation of some blockcipher modes of operation,''
  \emph{Cryptography Research and Evaluation Committees (CRYPTREC) for the
  Government of Japan}, 2011.

\bibitem{bellare2005secure}
M.~Bellare, T.~Kohno, and C.~Namprempre, ``The secure shell (ssh) transport
  layer encryption modes,'' Tech. Rep., 2005.

\bibitem{dworkin2001recommendation}
M.~Dworkin, ``Recommendation for block cipher modes of operation. methods and
  techniques,'' NATIONAL INST OF STANDARDS AND TECHNOLOGY GAITHERSBURG MD
  COMPUTER SECURITY DIV, Tech. Rep., 2001.

\bibitem{ylonen2005arch}
T.~Ylonen and C.~Lonvick, ``The secure shell (ssh) protocol architecture,''
  Tech. Rep., 2005.

\bibitem{ylonen2005transport}
------, ``The secure shell (ssh) transport layer protocol,'' Tech. Rep., 2005.

\bibitem{ylonen2005auth}
------, ``The secure shell (ssh) authentication protocol,'' Tech. Rep., 2005.

\bibitem{ylonen2005connection}
------, ``The secure shell (ssh) connection protocol,'' Tech. Rep., 2005.

\bibitem{barrett2001ssh}
D.~J. Barrett, D.~J. Barrett, R.~E. Silverman, and R.~Silverman, \emph{SSH, the
  Secure Shell: the definitive guide}.\hskip 1em plus 0.5em minus 0.4em\relax "
  O'Reilly Media, Inc.", 2001.

\bibitem{PuTTY}
S.~Tatham, ``{“PuTTY”},''
  \url{https://www.chiark.greenend.org.uk/~sgtatham/putty/latest.html}, 2019,
  [Online; accessed 29-Jan-2019].

\bibitem{galbraith2006ssh}
J.~Galbraith and O.~Saarenmaa, ``Ssh file transfer protocol,'' \emph{Work in
  Progress}, 2006.

\bibitem{vomel2013evaluation}
S.~V{\"o}mel and J.~St{\"u}ttgen, ``An evaluation platform for forensic memory
  acquisition software,'' \emph{Digital Investigation}, vol.~10, pp. S30--S40,
  2013.

\bibitem{vomel2011survey}
S.~V{\"o}mel and F.~C. Freiling, ``A survey of main memory acquisition and
  analysis techniques for the windows operating system,'' \emph{Digital
  Investigation}, vol.~8, no.~1, pp. 3--22, 2011.

\bibitem{FTKImager}
AccessData, ``{"FTK Imager"},''
  \url{http://marketing.accessdata.com/ftkimager4.2.0}, 2018, [Online; accessed
  29-Jan-2019].

\bibitem{Memoryze}
"FireEye", ``{“Memoryze”},''
  \url{https://www.fireeye.com/services/freeware.html}, 2018, [Online; accessed
  29-Jan-2019].

\bibitem{WinPMEM}
M.~Cohen", ``{“WinPMEM”},''
  \url{https://github.com/google/rekall/tree/master/tools/windows/winpmem},
  2018, [Online; accessed 29-Jan-2019].

\bibitem{freiling2018advances}
F.~Freiling, T.~Gro{\ss}, T.~Latzo, T.~M{\"u}ller, and R.~Palutke, ``Advances
  in forensic data acquisition,'' \emph{IEEE Design \& Test}, vol.~35, no.~5,
  pp. 63--74, 2018.

\bibitem{LiME}
J.~Sylve", ``{“LiME ~ Linux Memory Extractor”},''
  \url{https://github.com/504ensicslabs/lime}, 2019, [Online; accessed
  29-Jan-2019].

\bibitem{yang2016tool}
H.~Yang, J.~Zhuge, H.~Liu, and W.~Liu, ``A tool for volatile memory acquisition
  from android devices,'' in \emph{IFIP International Conference on Digital
  Forensics}.\hskip 1em plus 0.5em minus 0.4em\relax Springer, 2016, pp.
  365--378.

\bibitem{sun2015reliable}
H.~Sun, K.~Sun, Y.~Wang, and J.~Jing, ``Reliable and trustworthy memory
  acquisition on smartphones,'' \emph{IEEE Transactions on Information
  Forensics and Security}, vol.~10, no.~12, pp. 2547--2561, 2015.

\bibitem{Cellebrite}
"Cellebrite", ``{“Advanced Extraction Service”},''
  \url{https://www.cellebrite.com/en/services/advanced-extraction-services},
  2018, [Online; accessed 29-Jan-2019].

\bibitem{case2017memory}
A.~Case and G.~G. Richard~III, ``Memory forensics: The path forward,''
  \emph{Digital Investigation}, vol.~20, pp. 23--33, 2017.

\bibitem{yang2017live}
S.~J. Yang, J.~H. Choi, K.~B. Kim, R.~Bhatia, B.~Saltaformaggio, and D.~Xu,
  ``Live acquisition of main memory data from android smartphones and
  smartwatches,'' \emph{Digital Investigation}, vol.~23, pp. 50--62, 2017.

\bibitem{kondapally2018}
B.~P. Kondapally", ``{“What is IoT Forensics and How is it Different from
  Digital Forensics?}”,''
  \url{https://securitycommunity.tcs.com/infosecsoapbox/articles/2018/02/27/what-iot-forensics-and-how-it-different-digital-forensic},
  2018, [Online; accessed 29-Jan-2019].

\bibitem{alabdulsalam2018internet}
S.~Alabdulsalam, K.~Schaefer, T.~Kechadi, and N.-A. Le-Khac, ``Internet of
  things forensics: Challenges and case study,'' \emph{arXiv preprint
  arXiv:1801.10391}, 2018.

\bibitem{LibVMI}
{"LibVMI Project"}, ``{“LibVMI”},'' \url{http://libvmi.com/}, 2013,
  [Online; accessed 29-Jan-2019].

\bibitem{pyvmi}
B.~D. Payne", ``{“pyvmi -- A Python adapter for LibVMI”},''
  \url{https://github.com/libvmi/libvmi/tree/master/tools/pyvmi}, 2013,
  [Online; accessed 29-Jan-2019].

\bibitem{volatility2016}
{"The Volatility Foundation"}, ``{“The Volatility Foundation - Open Source
  Memory Forensics”},'' \url{http://www.volatilityfoundation.org/}, 2017,
  [Online; accessed 29-Jan-2019].

\bibitem{rekall2017}
M.~Cohen", ``{Rekall Memory Forensic Framework”},''
  \url{http://www.rekall-forensic.com/}, 2017, [Online; accessed 29-Jan-2019].

\bibitem{shannon1948mathematical}
C.~E. Shannon, ``A mathematical theory of communication,'' \emph{Bell system
  technical journal}, vol.~27, no.~3, pp. 379--423, 1948.

\bibitem{cachin1997entropy}
C.~Cachin, ``Entropy measures and unconditional security in cryptography,''
  Ph.D. dissertation, ETH Zurich, 1997.

\bibitem{xen2018}
{"Xen Project"}, ``{Xen Project Software Overview”},''
  \url{https://wiki.xenproject.org}, 2018, [Online; accessed 27-Nov-2018].

\bibitem{netfilterqueue}
{Kerkhoff Technologies}, ``{NetFilterQueue},''
  \url{https://pypi.org/project/NetfilterQueue}, 2017, [Online; accessed
  29-Jan-2019].

\bibitem{scapy2017}
P.~Biondi", ``{Scapy”},'' \url{https://scapy.readthedocs.io/en/latest/},
  2017, [Online; accessed 29-Aug-2018].

\bibitem{pycrypto2018}
D.~C. Litzenberger", ``{Python Cryptography Toolkit (pycrypto)”},''
  \url{http://www.rekall-forensic.com/}, 2013, [Online; accessed 29-Jan-2018].

\bibitem{ssh2018}
{"SSH Communications"}, ``{SSH Client for Windows - Comparison”},'' \url{
  https://www.ssh.com/ssh/client}, 2018, [Online; accessed 29-Jan-2018].

\bibitem{ylonen2015security}
T.~Ylonen, P.~Turner, K.~Scarfone, and M.~Souppaya, ``Security of interactive
  and automated access management using secure shell (ssh),'' Tech. Rep., 2015.

\bibitem{patel2017}
B.~Patel", ``{Intel Releases New Technology Specification for Memory
  Encryption”},''
  \url{https://software.intel.com/en-us/blogs/2017/12/22/intel-releases-new-technology-specification-for-memory-encryptio},
  2017, [Online; accessed 15-Oct-2018].

\bibitem{nichols2018}
S.~Nichols", ``{Epyc fail? We can defeat AMD’s virtual machine encryption,
  say boffins,”},'' \url{https://www.theregister.co.uk}, 2017, [Online;
  accessed 15-Oct-2018].

\end{thebibliography}

\end{document}